\documentclass[journal]{IEEEtran}

\usepackage[english]{babel}
\usepackage{mathptmx}
\usepackage{array}
\usepackage{multirow}
\usepackage{tabularx}
\usepackage{amsmath}
\usepackage{graphicx}
\usepackage[colorlinks=true, allcolors=blue]{hyperref}
\usepackage{float}
\usepackage[normalem]{ulem}
\usepackage{cite}
\usepackage{amsfonts}

\begin{document}

\title{Bending beams behind corners: mechanisms, challenges and capabilities for wireless connectivity}

\author{{\IEEEauthorblockN{Sotiris Droulias and Angeliki Alexiou,~\IEEEmembership{Member,~IEEE}} }
\thanks{The authors are with the Department of Digital Systems, University of Piraeus, Piraeus 18534, Greece (Corresponding author: Sotiris Droulias, e-mail: sdroulias@unipi.gr)}}
\maketitle
\begin{abstract}
Curved beams, that is, beams that are able to propagate on nonlinear trajectories, are often envisioned as ideal candidates for blockage avoidance in future wireless connectivity. Owing to this unique feature, they are considered as ideal beams for bending around and behind corners to reach users beyond the line-of-sight (LoS), thus offering unprecedented connectivity. In this work, we explain the various mechanisms of beam propagation beyond the LoS, and we demonstrate that beam bending behind corners results from an interplay between wavefront engineering and edge diffraction, with distinct characteristics that depend on the extent of blockage and the beam formation efficiency. We identify three distinct regimes of operation, namely the unblocked, the partially blocked, and the fully blocked regime, and we show that beam bending through wavefront engineering dominates in the unblocked and partially blocked regimes, while edge diffraction dominates in the fully blocked regime; as a result, curved beams cannot really bend behind the corner, unless there is some LoS between the user and the transmitter. Based on our findings, we compare curved beams with focused beams, and we demonstrate that they perform similarly in the partially blocked regime, while focused beams outperform curved beams in the unblocked and fully blocked regimes.
\end{abstract}


\section{Introduction}
\IEEEPARstart{A}{s} wireless communications are nowadays shifting from the far-field to the near-field of antenna systems, the concept of wavefront engineering for applications in near-field wireless connectivity is gradually gaining more ground \cite{Stratidakis2024}. Tailoring the beam's wavefront to acquire curvature beyond the typical far-field planar form used in conventional beam-forming, opens up opportunities for generating new types of beams that can focus power at selected areas \cite{Eldar2022, Dai2023, Droulias2024a}, propagate without diffracting \cite{Droulias2024b,Mutai2025} or even evolve on bent trajectories \cite{Droulias2025}. The latter are commonly referred to as \textit{curved beams}.   \\
\indent While in previous years the concept of wavefront engineering for generating curved beams has flourished mainly in the realm of optics, recently there has been an extensive effort to realize such beams in the GHz and low THz frequencies \cite{Qu2023,Zhang2024,Hu2025}, with the aim to transfer their intriguing functionalities in the wireless domain. For example, owing to their unique property of circumventing undesired objects or users, curved beams have been studied in the context of blockage avoidance \cite{Guerboukha2024,Chen2025,Zhao2026,Qin2026}. Recent studies have focused on different optimization approaches \cite{Liu2025,Uchimura2025,Uchimura2026a,Uchimura2026b}, for designing curved beams that meet specific performance metrics. As curved beams could circumvent a potential eavesdropper that attempts to tap the communication link, they have also been studied for enhanced physical layer security over conventional beamforming \cite{Petrov2024,Petrov2025,Stults2026,Droulias2026}. 
Very recently, ways to generate curved beams with apertures that are not necessarily flat have been investigated \cite{Canals2025}, offering increased LoS between the transmitter and targeted users or areas of interest. \\
\indent Curved beams can undoubtedly bypass blockers, as has been demonstrated both theoretically and experimentally in several works. Yet, could they also bend entirely behind them? Such a possibility would be crucial at high frequencies, where multipath is poor and blockers cast a shadow to users that reside behind them. Importantly, are curved beams the optimal choice for such scenarios or are there some other beam profiles that might be preferable? Although some works demonstrate that curved beams are the optimal choice \cite{Guerboukha2024,Zhao2026,Qin2026}, and suggest that bending behind the blocker is possible, there are several crucial aspects that need to be taken into account to fully understand the underlying mechanisms of beam bending around and behind a blocker. 
One such aspect is the beam formation efficiency, which depends on the electrical size of the transmitter, i.e. its actual size relative to the operating wavelength. Another aspect is the extent of blockage, that is, how much of the LoS is blocked. Last, there is an interplay between wavefront engineering and edge diffraction, which is often overlooked \cite{Pallaprolu2025}. Understanding how these aspects determine beam bending is crucial for understanding under which circumstances curved beams are indeed the ideal choice or whether there are other effective alternatives.\\
\indent To address these aspects, in this work, we analyze the mechanisms of beam propagation beyond the LoS. We demonstrate that beam bending in the presence of a blocker results from the interplay between wavefront engineering and edge diffraction, with distinct characteristics that depend on the extent of blockage and the beam formation efficiency. The main contributions are summarized as follows. 
\begin{itemize}
    \item We explain the three basic mechanisms of beam propagation beyond the LoS, namely refractive index modulation, edge diffraction and wavefront engineering.
    \item We demonstrate that beam bending around and behind the corner results from the interplay between wavefront engineering and edge diffraction.
    \item We derive analytical expressions to quantify and assess the efficiency of wavefront engineering and edge diffraction as a function of the operating frequency. We study curved beams, which are the natural choice for bent propagation, and we compare them with focused beams.
    \item We identify three distinct regimes of operation, namely the unblocked, partially blocked, and fully blocked, and we show that propagation beyond the LoS through wavefront engineering dominates in the unblocked and partially blocked regimes, while edge diffraction dominates in the fully blocked regime.
    \item By performing full parametric scan we optimize the curved beams and the focused beams, to maximize the power at users located in all three regimes. Our findings reveal that optimized focused beams can generally perform at least as efficiently as optimized curved beams in beyond-the-LoS propagation scenarios.
\end{itemize}
%

%
%
\begin{figure}[t!]
\centering
    \includegraphics[width=0.95\linewidth]{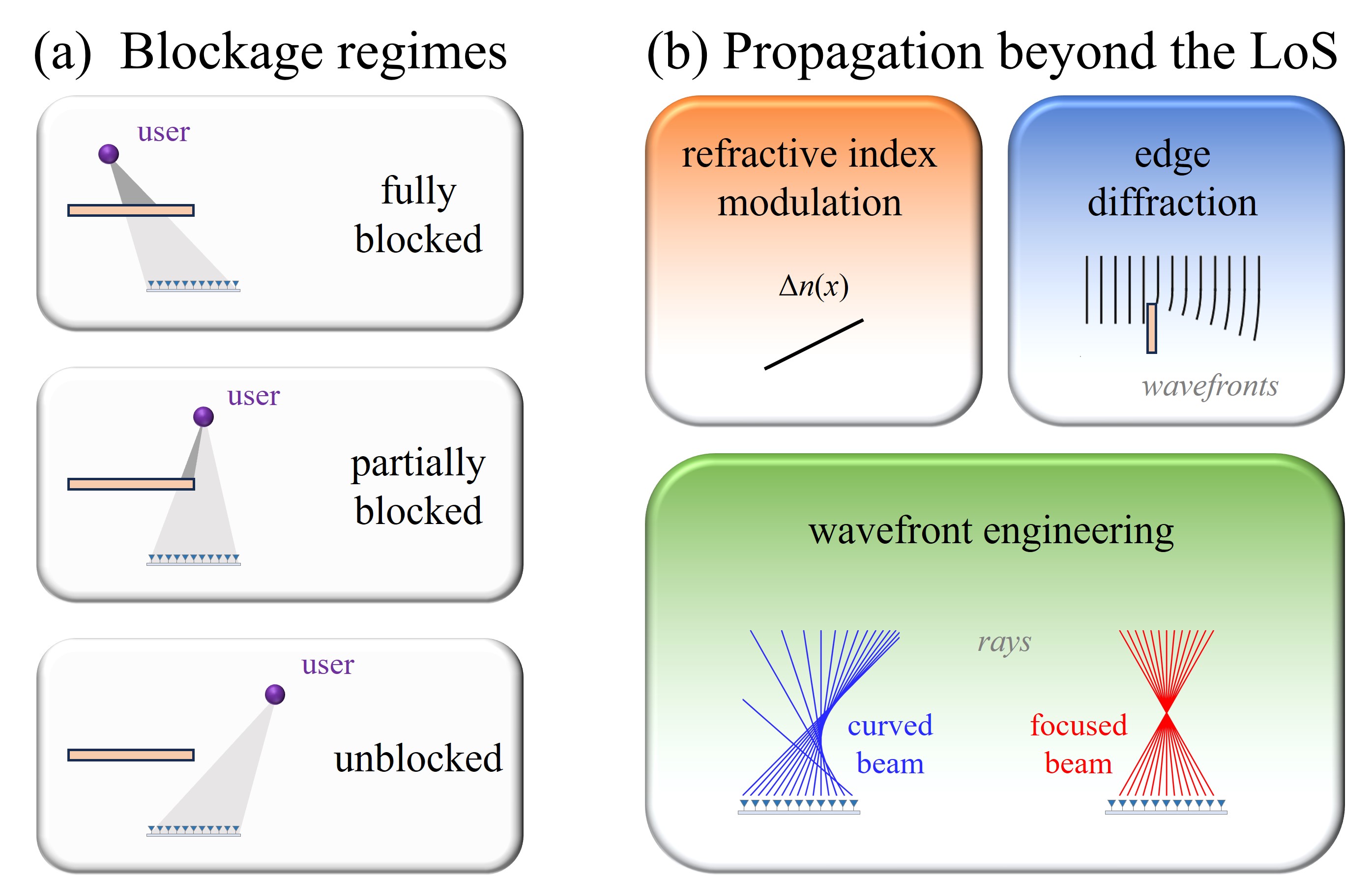}    
	\caption{Schematic representation of (a) blockage regimes and (b) mechanisms of beam propagation beyond the LoS.}    
    	\label{fig:fig01}
\end{figure}
%
%


%
\section{Mechanisms of beam bending}
\noindent We consider a communication link scenario, in which the transmitter (Tx) is located at the origin of the coordinate system. The Tx is equipped with a uniform planar array (UPA) of size $L_x\times L_y$ extending on the $xy$-plane, which generates beams that propagate along the $z$-direction. Each beam is directed towards the user, and is possibly blocked. Depending on the extent of the beam that is blocked, we identify three blockage regimes, namely fully blocked, partially blocked, and unblocked, as shown schematically in Fig.\,\ref{fig:fig01}(a). \\
\indent To study beam propagation in free-space, we start from Maxwell's equations, which in a homogeneous medium reduce to the Helmholtz equation ($ \nabla^2+k^2) \textbf{\textrm E}(\textbf{\textrm r})=0$, where $\nabla^2$ is the Laplacian operator, $\textbf{\textrm E}(\textbf{\textrm r})$ is the vector of the electric field at position $\textbf{\textrm r}=(x,y,z)$ and $k=2\pi/\lambda$ is the wavenumber ($\lambda$ is the wavelength). An identical equation holds for the $\textbf{\textrm H}$-field. The Helmoltz equation defines three independent scalar equations, namely one for $E_x$, one for $E_y$, and one for $E_z$. 
For simplicity, we consider beams that are invariant along the $y$-direction, to which we refer as \textit{one-dimensional} (1D). For such beams $L_y\rightarrow \infty$, in practice referring to UPAs with $L_y\gg L_x$. Our results are directly extended to 2D beams, for which $L_x,L_y$ are of the same order of magnitude. For 1D $y$-polarized paraxial waves ($E_y\equiv E(x,z)$), the Helmholtz equation takes the form 
\begin{align}
    j\frac{\partial E}{\partial z} + \frac{1}{2kn_0}\frac{\partial^2 E}{\partial x^2} + k\Delta n(x)E = 0,
    \label{Eq:EqHelmholtz}
\end{align}
where $n_0$ is the refractive index (RI) of the homogeneous medium and $\Delta n(x)$ accounts for possible RI modulation. For example, in vacuum $n_0=1$ and $\Delta n(x)=0$. The beam profile at the input plane is a function of $x$ and has the general form
\begin{align}
    E(x,z=0)=A(x) e^{j\phi(x)},
    \label{Eq:EqINPUTWAVE}
\end{align}
where $A$ is the amplitude and $\phi$ the phase. Because beam formation is determined primarily by the input phase $\phi$ \cite{Droulias2025}, the amplitude $A$ across all UPA elements is taken as constant, $A(x)=A_0$, simplifying the excitation conditions in realistic systems. In view of \eqref{Eq:EqHelmholtz} and \eqref{Eq:EqINPUTWAVE}, we identify three major mechanisms, which are schematically depicted in Fig.\,\ref{fig:fig01}(b).

%
%
\begin{figure}[t!]
\centering
    \includegraphics[width=1\linewidth]{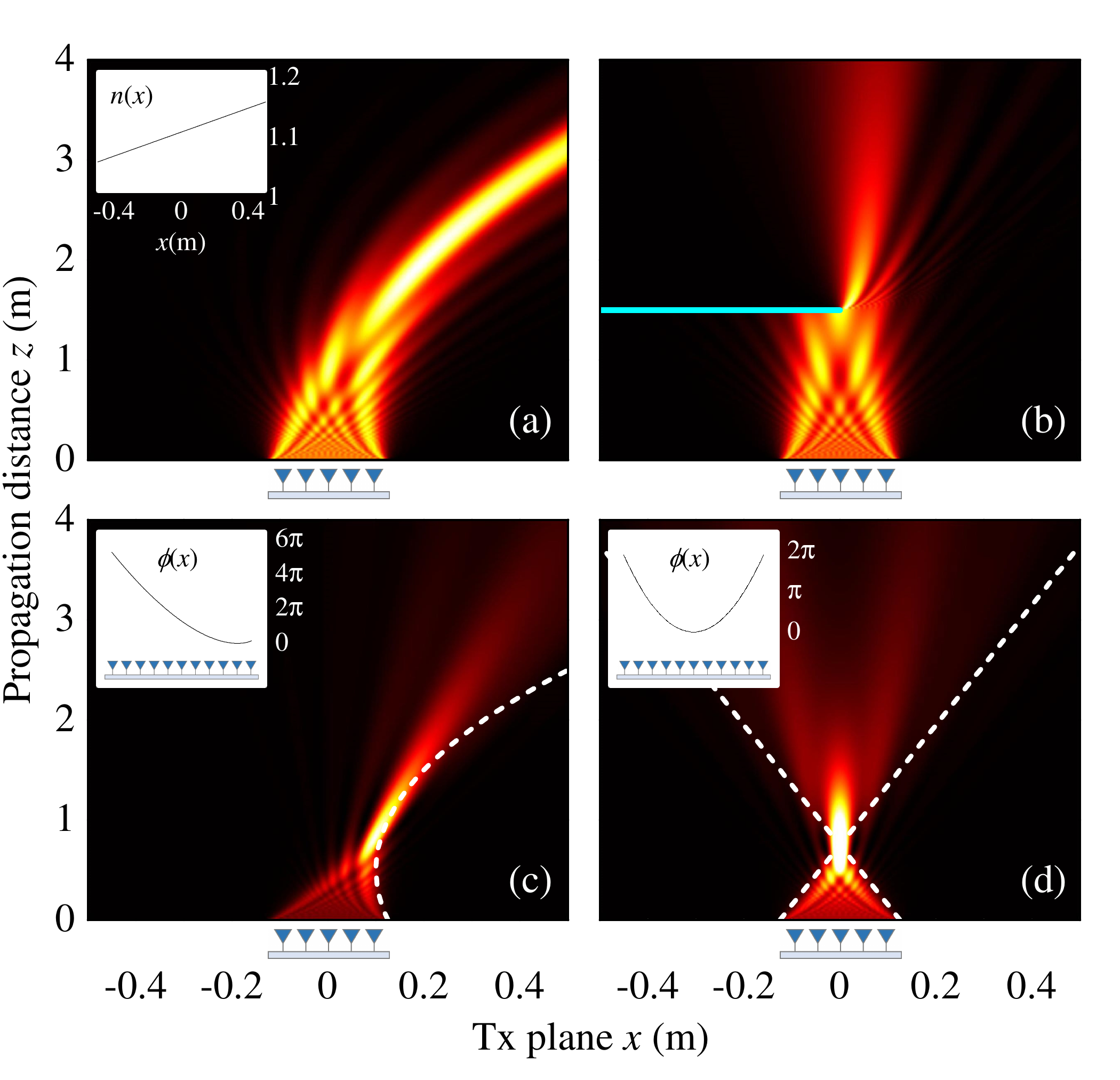}    
	\caption{Demonstration of beam propagation mechanisms beyond the LoS, using a UPA with size $L_x=0.25\,\mathrm{m}$ ($L_y\gg L_x$) at $28\,\mathrm{GHz}$. (a) Refractive index modulation. (b) Knife edge diffraction. (c) Curved beam. (d) Focused beam.}
    	\label{fig:fig02}
\end{figure}
%

\subsection{RI modulation: tailoring the medium}

A straightforward way to guide the beam on a bent path is by engineering the medium in which the beam propagates. This can be achieved by creating a gradient potential in \eqref{Eq:EqHelmholtz} by means of $\Delta n(x)$. One such example is demonstrated in Fig.\,\ref{fig:fig02}(a), where $n_0=1$ and $\Delta n(x) = 0.1(x+1)$, $|x|<1$, using a UPA with $L_x=0.25\,\mathrm{m}$ and operating frequency $f=28\,\mathrm{GHz}$. This approach essentially mimics the origin of the well-known mirage effect, which is caused by light bending in layers of air with different RI, due to varying temperature. In this example, the RI change has been intentionally considered to be much stronger than what is typically feasible in air, the RI of which depends weakly on the temperature \cite{Owens1967}. Hence, in realistic scenarios, such a possibility would require much longer propagation distances.
For the numerical propagation of the beam, here and in all examples in this work, we have used the angular spectrum representation \cite{Droulias2022, Droulias2024a}, which implements the exact scalar Helmholtz propagation on a sampled grid, including evanescent components (see Appendix \ref{Sec:AppendixA} for details).

\subsection{Edge diffraction}

Another possibility is to take advantage of diffraction, i.e. set $\Delta n(x)=0$ and introduce a knife edge along the beam propagation path. One such example is demonstrated in Fig.\,\ref{fig:fig02}(b), where the same beam used in the previous example encounters a blocker of infinite extent along the $y$-direction after $1.5\,\mathrm{m}$ of propagation. The presence of the blocker causes some part of the beam to bend behind it, due to diffraction from the edge of the blocker.

\subsection{Wavefront curvature engineering}

Lastly, instead of tailoring the environment, one can engineer the input beam profile so that constructive interference occurs along a bent path. This can be achieved by judicious design of the amplitude and phase in \eqref{Eq:EqINPUTWAVE}, which is the initial condition for \eqref{Eq:EqHelmholtz}. Because in this work the amplitude $A$ is uniform for all UPA elements, we will focus on the design of the phase $\phi(x)$.

\subsubsection{Curved beams}

The underlying mechanism of how curved beams propagate on bent trajectories can be understood in terms of their equivalent rays \cite{Droulias2025}. While rays necessarily evolve along straight paths, and thus cannot bend, the desired bending results from their envelope, namely their \textit{caustic}, which corresponds to the trajectory of the beam's main lobe. To form the desired caustic, one needs to engineer the slope of the rays at their origin, i.e. the input phase $\phi$, which is associated with the wavefront curvature of the wave. \\
\indent For curved beams with their main lobe propagating along the parabolic trajectory $x=x_0+\beta (z-z_0)^2$ with vertex located at $(x_0,z_0)$, the necessary phase is given by \cite{Droulias2025}
\begin{align}    
    \phi(x) = 2\beta k z_0 x + \frac{4}{3} \beta^2 k \left(z_0^2 + \frac{x_0-x}{\beta} \right)^\frac{3}{2},
    \label{Eq:EqAiryPHASE}
\end{align}
which is real-valued for $x<x_0+\beta z_0^2\equiv x_{0C}$. Note that $x_{0C}$ marks the location of the trajectory at $z=0$. In Fig.\,\ref{fig:fig02}(c) we demonstrate one such beam, for which $x_0=0.1\,\mathrm{m}$, $z_0=0.5\,\mathrm{m}$ and $\beta=0.1\,\mathrm{m^{-1}}$. The dashed line marks the designed parabolic trajectory.

\subsubsection{Focused beams}

Beams that are able to concentrate the transmitted power onto a confined region in space, the so-called \textit{focal point}, rapidly converge towards it and diverge beyond it. 
As such, they may offer an alternative to beam propagation beyond the LoS, if a blocker is located close to the focal point. The underlying mechanism of how beams are focused can be understood in terms of their equivalent rays:  they are all targeted towards the focal point. \\
\indent Because the rays are all normal to concentric spheres that converge to the focal point, the necessary phase for beam focusing is such that a converging spherical wavefront is formed \cite{Droulias2024a}. In the 1D case, the beam wavefront reduces to cylindrical, and the necessary phase is therefore given by 
\begin{align}    
    \phi(x) = k \sqrt{\left(x-d_f\sin{\theta_r} \right)^2 + \left(d_f\cos{\theta_r}\right)^2},
    \label{Eq:EqPHASEfocus}
\end{align}
where $d_f$ is the focal distance and $\theta_r$ is the steering angle. In Fig.\,\ref{fig:fig02}(d) we demonstrate one such beam, for which $d_f=0.75\,\mathrm{m}$, $\theta_r=0^\circ$. The dashed lines that start from the edges of the UPA mark the outer rays, in which the major part of the beam is contained.
\section{Bending beams behind corners}
\noindent The technique of RI modulation is usually applied in dense media, like dielectrics, where strong RI modulation is accessible. Characteristic examples are optical fibers, waveguide arrays and graded index waveguides. In free-space, while RI modulation could potentially provide propagation beyond the LoS with a controllable radius of curvature, given the weak temperature-dependence of the air RI, such a possibility is rather challenging. 
In the remainder of this paper we will focus on the two other mechanisms, namely, edge diffraction and wavefront engineering. In this section, we derive analytical expressions to quantify and assess their efficiency with respect to the operating frequency across a wide spectral range from the FR2 up to the sub-THz.
\subsection{From wave optics to the ray optics limit}
To study the efficiency of the various mechanisms, we will analytically calculate the power density distribution in space for each case. Using the Huygens-Fresnel integral under the Fresnel approximation, the $E$-field at the observation point $\mathbf{r}=(x,y,z)$ is expressed as (see Appendix \ref{Sec:AppendixB} for details)
\begin{align}
    \label{Eq:EqHFintegral}
    & E(x,y,z) =-\frac{ik e^{-ik z}}{2\pi z} \times \\
    & \int_{-L_y/2}^{+L_y/2} \int_{-L_x/2}^{+L_x/2} A_0e^{j\phi(x')}e^{\frac{-ik}{2z}\left[(x-x')^2+(y-y')^2\right] } dx' dy',
    \nonumber
\end{align}
where $A_0$ is the uniform amplitude, $\phi(x)$ is the phase, and integration takes place across the UPA surface. For total input beam power $P_t$, i.e. $\iint |E|^2/2\eta=P_t$ at $z=0$, where $\eta$ is the free-space wave impedance, the amplitude is $A_0=\sqrt{2\eta P_t/L_xL_y}$, or $A_0=\sqrt{2\eta P_t/L_x}$ for $L_y\rightarrow \infty$.

\subsubsection{Edge diffraction}
For conventional beam-forming, i.e. with no wavefront engineering, we set $\phi(x)=0$. Application of \eqref{Eq:EqHFintegral} in this case leads to 
\begin{align}
    \label{Eq:EqEdiff}
    & E(x,y,z) =  -A_0e^{-ikz}\\ 
    & \frac{\mathrm{erf} (q_{x+})+\mathrm{erf}(q_{x-})}{2} \,
    \frac{\mathrm{erf} (q_{y+})+\mathrm{erf}(q_{y-})}{2},
    \nonumber
\end{align}
where $\mathrm{erf(\cdot)}$ is the error function, and $q_{x\pm},q_{y\pm}$ are given by
\begin{align}
    \label{Eq:EqA1qx}
    q_{x\pm} = \left(\frac{1}{4}+\frac{i}{4} \right) \sqrt{\frac{k}{z}}(L_x\pm2 x), \\
    \label{Eq:EqA1qy}
    q_{y\pm} = \left(\frac{1}{4}+\frac{i}{4} \right) \sqrt{\frac{k}{z}}(L_y\pm2 y).
\end{align}
This is the field that is generated by the UPA, and is equivalent to a beam emerging at $z=0$ from an aperture of size $L_x\times L_y$ that creates an opening in an otherwise opaque blocker of infinite extent in the $xy$-plane. Therefore, at a certain observation point with $z>0$, $|x|>L_x/2$, and $|y|>L_y/2$, the field resides entirely in the shadow of that blocker.
Using \eqref{Eq:EqEdiff}, the power density $S=|E|^2/2\eta$ is
\begin{align}
    \label{Eq:EqSdiff}
    S_\mathrm{diff} =  \frac{A_0^2}{2\eta}
    \left| \frac{\mathrm{erf} (q_{x+})+\mathrm{erf}(q_{x-})}{2} \right|^2
    \left| \frac{\mathrm{erf} (q_{y+})+\mathrm{erf}(q_{y-})}{2} \right|^2,
\end{align}
where the subscript denotes \textit{diffraction}. For 1D beams $L_y\rightarrow \infty$, which leads to $[\mathrm{erf} (q_{y+})+\mathrm{erf}(q_{y-})]/2 \rightarrow 1$. For $k\gg (L_x\pm x)^2/z$, we can approximate the remaining $\mathrm{erf} (q_{x\pm})$ terms, so that the power density becomes
\begin{align}
    \label{Eq:EqSdiffapprx}
    S_\mathrm{diff,1D} = \frac{A_0^2}{2\eta} \frac{4z(L_x^2+4 x^2)}{\pi(L_x^2-4 x^2)^2}\frac{1}{k} \left(1+ \frac{L_x^2-4 x^2}{L_x^2+4 x^2} \cos{\left(\frac{kL_x x}{z}\right)} \right).
\end{align}
The power density is the sum of an oscillatory term and a term that is proportional to $\sim 1/k$. Eliminating the oscillatory term, leads to
\begin{align}
    S_\mathrm{diff,1D} = \frac{cP_t}{2\pi} \frac{4z(L_x^2+4 x^2)}{\pi L_x(L_x^2-4 x^2)^2}\frac{1}{f},
    \label{Eq:EqSdiff1D}
\end{align}
where $c$ is the vacuum light speed and we have substituted the uniform amplitude for 1D beams, $A_0=\sqrt{2\eta P_t/L_x}$ (see Appendix \ref{Sec:AppendixC} for details on the approximations). Hence, $S_\mathrm{diff,1D}\sim 1/f$. The reduction of the field strength with increasing frequency is not unexpected; it simply expresses the transition to the ray optics limit, where the shadow behind a blocker becomes sharper.
Repeating the calculation for 2D beams, leads to 
\begin{align}
    S_\mathrm{diff,2D} = \frac{c^2P_t}{(2\pi)^2} \frac{4z(L_x^2+4 x^2)}{\pi L_x(L_x^2-4 x^2)^2}\frac{4z(L_y^2+4 y^2)}{\pi L_y(L_y^2-4 y^2)^2}\frac{1}{f^2},
    \label{Eq:EqSdiff2D}
\end{align}
i.e. $S_\mathrm{diff,2D}\sim f^{-2}$, where we have used the uniform amplitude for 2D beams, $A_0=\sqrt{2\eta P_t/L_xL_y}$. Hence, the power density is generally expressed as $S_\mathrm{diff,nD}\sim f^{-n}$, with $n=\{1,2\}$.

\subsubsection{Focused beams}
\indent For 1D focused beams, we use \eqref{Eq:EqPHASEfocus}, which for $\theta_r=0^\circ$ becomes $\phi(x)=k \sqrt{x^2 + d_f^2}$. To perform the integrations in \eqref{Eq:EqHFintegral} we simplify the phase using the parabolic approximation \cite{Droulias2024a} $\phi(x)=kd_f+k x^2/2d_f$. The calculation of the Huygens-Fresnel integral leads to the result
\begin{align}
    \label{Eq:EqEfoc}
    & E(x,y,z) =  - \frac{A_0}{\sqrt{1-\frac{z}{d_f}}} e^{-ik(z-d_f+\frac{x^2}{2(z-d_f)})} \\ 
    &  \frac{\mathrm{erf} (q_{x+}p_+)+\mathrm{erf}(q_{x-}p_-)}{2}  \,
    \frac{\mathrm{erf} (q_{y+})+\mathrm{erf}(q_{y-})}{2} ,
    \nonumber
\end{align}
where $q_{x\pm},q_{y\pm}$ are given by \eqref{Eq:EqA1qx},\eqref{Eq:EqA1qy}, and 
\begin{align}
    p_{\pm} = \frac{1}{\sqrt{1-\frac{z}{d_f}}} \left(1-\frac{L_x z}{d_f(L_x\pm2 x)}  \right).
\end{align}
Using \eqref{Eq:EqEfoc}, the power density $S=|E|^2/2\eta$ at the focal point ($x=0,y=0,z=d_f$) is
\begin{align}
    S_\mathrm{foc,1D} = \frac{1}{2\eta}\frac{A_0^2 L_x^2}{2\pi d_f}k= \frac{P_t L_x}{cd_f}f,
    \label{Eq:EqSfoc1D}
\end{align}
where the subscript denotes 1D-\textit{focusing}, and we have used $A_0=\sqrt{2\eta P_t/L_x}$. Hence, $S_\mathrm{foc,1D}\sim f$.
Repeating the calculation for 2D focusing, i.e. using $\phi(x)= kd_f + k (x^2+y^2)/2d_f$ and $A_0=\sqrt{2\eta P_t/L_xL_y}$, leads to
\begin{align}
    S_\mathrm{foc,2D} = \frac{1}{2\eta}\frac{A_0^2 L_x^2L_y^2}{(2\pi)^2 d_f^2}k^2= \frac{P_t L_xL_y}{c^2 d_f^2}f^2,
    \label{Eq:EqSfoc2D}
\end{align}
i.e. $S_\mathrm{foc,2D}\sim f^2$. Hence, the power density at the focal point is generally expressed as $S_\mathrm{foc,nD}\sim f^n$, with $n=\{1,2\}$. \\
\subsubsection{Curved beams}
\indent For 1D curved beams, we can use the analytical form of the 1D Airy beam \cite{Droulias2025}, which evolves on the same parabolic path as the curved beams in this work. The $E$-field of the 1D Airy beam is
\begin{align}
    E(x,z)=E_0 A_i\left[(4\beta k^2)^{1/3} (x-\beta z^2+j\frac{\alpha}{k}z)\right]e^{j\psi(x,z)},
    \label{Eq:EqAiryTHEORY}
\end{align}
where $\psi = \left[2\beta k(x-(2/3)\beta z^2) + \alpha^2/2k\right] -j \alpha(x-2\beta z^2)$, $A_i$ is the Airy function and the parameters $\alpha$ and $\beta$ are real constants with units of $\mathrm{m}^{-1}$. $E_0$ is a complex constant, which for a total beam power $P_t$ is
\begin{align}
    E_0 = \sqrt{2\eta P_t\sqrt{8\pi\alpha (4\beta)^{1/3}}e^{-\frac{2}{3}\frac{\alpha^3}{4\beta k^2}}k^{1/3}}.
    \label{Eq:EqAiryTHEORY_E0}    
\end{align}
Using \eqref{Eq:EqAiryTHEORY}, the power density of the Airy beam $S=|E|^2/2\eta$ at the vertex of its parabolic trajectory ($x=0,z=0$) is 
\begin{align}
    \label{Eq:EqSAiry1D_at0}
    S_\mathrm{Airy,1D}=P_t\sqrt{8\pi\alpha (4\beta)^{1/3}}e^{-\frac{2}{3}\frac{\alpha^3}{4\beta k^2}}A_i(0)^2  k^{1/3},
\end{align}
where $A_i(0)^2\approx 0.126$. Because in our examples $\alpha^3/4\beta k^2\ll 1$, the exponential term can be eliminated, and we reach the result
\begin{align}
    \label{Eq:EqSAiry1D}
    S_\mathrm{Airy,1D}= P_t \left(\frac{2\alpha^3 \beta}{\pi c^2} \right)^\frac{1}{6} 4\pi A_i(0)^2 f^{1/3}.
\end{align}
Hence,  $S_\mathrm{Airy,1D}\sim f^{1/3}$. Table~\ref{tab1} summarizes the frequency dependence of the field strength $S$ for all mechanisms.

\renewcommand{\arraystretch}{1.8}

\begin{table}[b]
\begin{center}
\caption{Frequency dependence of field strength S \\ for main mechanisms of propagation beyond the LoS}
\label{tab1}

\begin{tabular}{|c|c|c|}
\hline \hline
\multirow{2}{*}{Edge diffraction} & \multicolumn{2}{c|}{Wavefront engineering} \\ \cline{2-3}
                   & Curved beam & Focused beam \\ \hline
$\sim f^{-n},\,\,\,n=\{1,2\}$      & $\sim f^{1/3}$ & $\sim f^n,\,\,\,n=\{1,2\}$ \\ 
\hline \hline
\end{tabular}

\end{center}
\end{table}

%
%
\begin{figure}[t!]
\centering
    \includegraphics[width=1\linewidth]{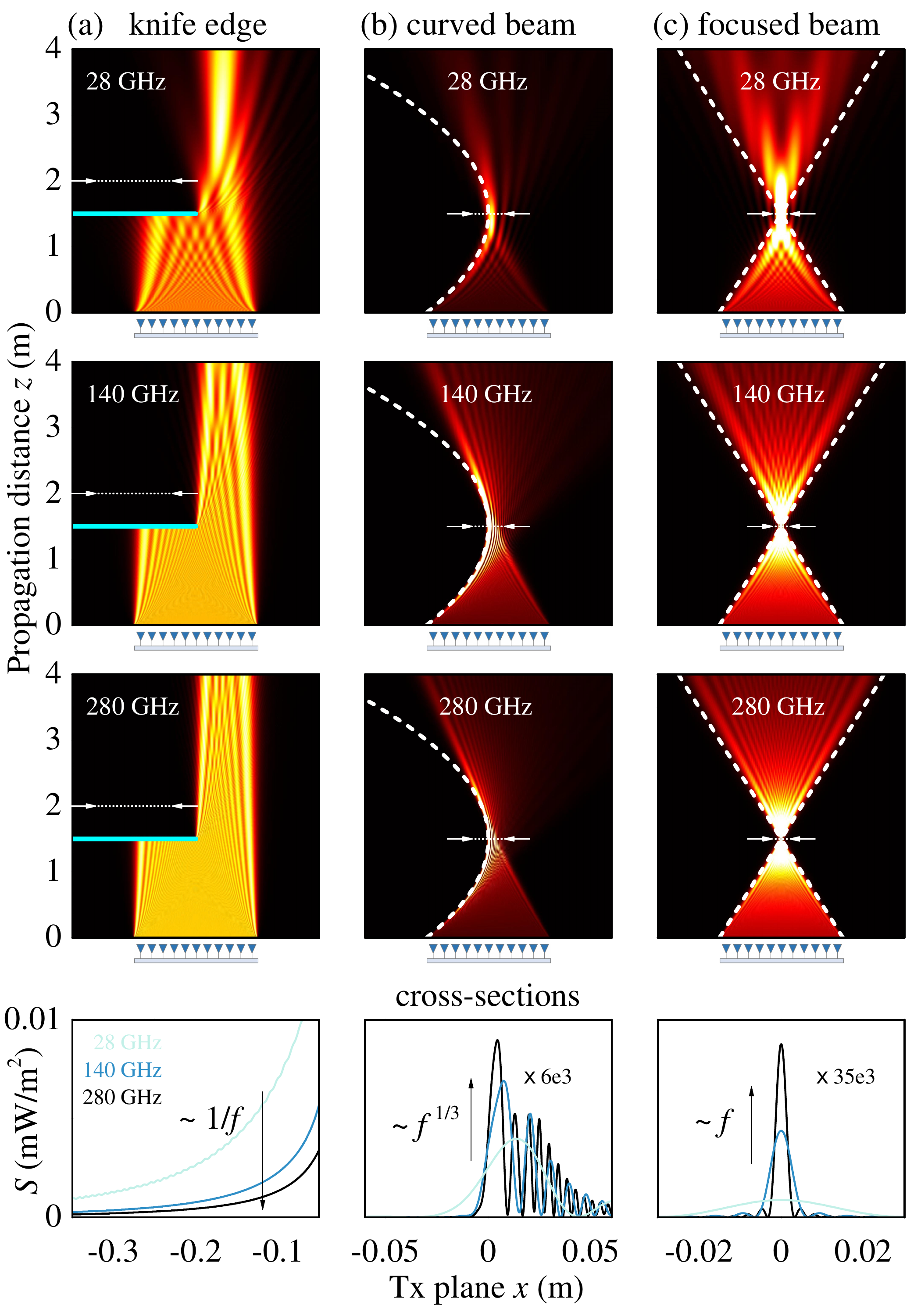}    
	\caption{Efficiency of beam propagation mechanisms beyond the LoS, as a function of the operating frequency, from the FR2 up to the sub-THz spectral range. (a) Knife edge diffraction. (b) Curved beam. (c) Focused beam. With increasing frequency, propagation beyond the LoS due to edge diffraction becomes less pronounced, while the efficiency of curved beam and focused beam formation improves. The arrows denote the cross-sections shown in the bottom panels.}
    	\label{fig:fig03}
\end{figure}
%

\subsubsection{Examples}

In Fig.\,\ref{fig:fig03} we demonstrate the efficiency of knife edge diffraction, curved beams and focused beams, at different operating frequencies. For a fair comparison we generate all 1D beams using the same aperture of extent $L_x=0.5\,\mathrm{m}$ and same total transmitted power $P_t=1\,\mathrm{mW}$. \\
\indent In Fig.\,\ref{fig:fig03}(a) we use a beam with $\phi(x)=0$, which at $z=1.5\,\mathrm{m}$ encounters a blocker of infinite extent along the $y$-direction, and is partly blocked. With increasing frequency, knife edge diffraction becomes less prominent, as a result of gradually shifting from the wave optics to the ray optics limit; the wavelength becomes smaller compared to the size of objects, creating sharper shadows behind blockers, with the field strength reducing as $\sim 1/f$, as analytically predicted by \eqref{Eq:EqSdiff1D}. \\
\indent On the other hand, curved beam and focused beam formation becomes more efficient towards the ray limit, as demonstrated in Fig.\,\ref{fig:fig03}(b) and Fig.\,\ref{fig:fig03}(c), respectively \cite{Droulias2025,Droulias2024a}. The curved beam is formed using \eqref{Eq:EqAiryPHASE} with  $x_0=0\,\mathrm{m}$, $z_0=1.5\,\mathrm{m}$, and $\beta=-0.11\,\mathrm{m^{-1}}$, and the focused beam is formed using \eqref{Eq:EqPHASEfocus} with $d_f=1.5\,\mathrm{m}$, $\theta_r=0^\circ$. In these examples, the field strength of the curved beam increases as $\sim f^{1/3}$, and of the focused beam as $\sim f$, as analytically predicted by \eqref{Eq:EqSfoc1D} and \eqref{Eq:EqSAiry1D}, respectively.

\subsection{Interplay between edge diffraction and wavefront engineering}

%
%
\begin{figure}[t!]
\centering
    \includegraphics[width=1\linewidth]{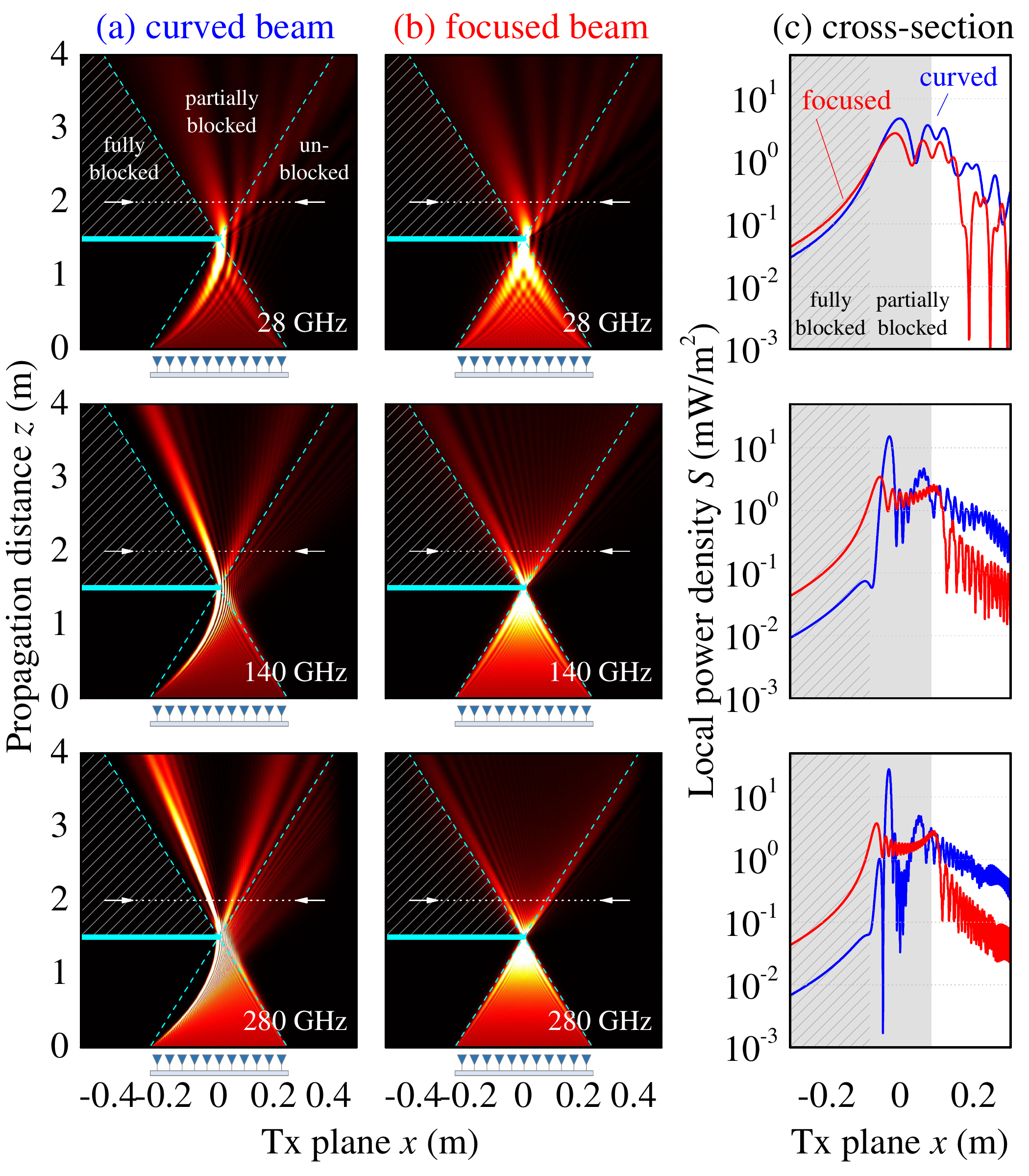}    
	\caption{Beam bending behind a corner, as an interplay between wavefront engineering and edge diffraction, from the FR2 up to the sub-THz spectral range. (a) Curved beam. (b) Focused beam. (c) Power density at the cross-section marked with the arrows in panels (a),(b).}
    	\label{fig:fig04}
\end{figure}
%
%

%
To maximize the power behind a blocker, one could intuitively bring the vertex of the curved beam and the focal point of the focused beam close to the edge of the blocker. Exactly at this location, a small part of the beam is blocked by the edge of the blocker and undergoes edge diffraction. Here we explore this scenario, to illustrate the interplay between edge diffraction and the individual wavefront engineering mechanisms.\\
\indent In Fig.\,\ref{fig:fig04} we use the curved beam and the focused beam of Figs.\,\ref{fig:fig03}(b),(c), with the blocker restored. The vertex of the curved beam and the focal point of the focused beam are designed to be co-located at $(x=0,z=1.5\,\mathrm{m})$, so that both beams intentionally `touch' the blocker's edge. As a result, they undergo some edge diffraction, as can be visually identified in Figs.\,\ref{fig:fig04}(a),(b). 
The stripes mark the area outlined by the UPA-blocker edge LoS, which determines the shadow region behind the blocker. This is the region where the UPA is fully blocked by the blocker. The two cyan dashed lines determine the region where the UPA is partially blocked by the blocker, i.e. there is some LoS between the UPA and a user that resides in this area; for any user located beyond these two regions, the UPA is unblocked, i.e. the user has LoS with the entire UPA. As can be observed, beam bending takes place predominantly in the partially blocked and unblocked regions. Yet, some fields are distributed in the fully blocked region, as can be identified in the local power density plots in Fig.\,\ref{fig:fig04}(c), which are extracted from the actual beams at the cross-section, marked with the arrows. These fields are dominated by edge diffraction rather than wavefront engineering, as can be verified in their frequency dependence in the shadow region; in the case of focused beam formation, the decrease of the local power density due to edge diffraction, which changes as $\sim f^{-1}$, is balanced by the increasing beam power at the blocker edge for 1D focusing, which changes as $\sim f$, resulting in practically constant field behind the blocker with respect to the operating frequency. However, in the case of curved beam formation, the beam power at the blocker edge increases at a slower rate as $\sim f^{1/3}$, resulting in an overall decrease in the field behind the blocker with respect to the operating frequency. Note that, because the power density of both beams at the blocker edge increases with frequency, pure beam bending behind the corner would manifest as power increase. Therefore, beam bending in the fully blocked region is dominated by edge diffraction rather than wavefront engineering, i.e. the beams cannot bend around corners per se. \\
\indent The results of Fig.\,\ref{fig:fig04} also reveal that focused beams can outperform curved beams in the fully blocked region. In the partially blocked region, both beams seem suitable, depending on the exact targeted location. Of course, because these beams are not optimized, they are not necessarily the most suitable choices for this particular configuration. Therefore, to validate the generality of our findings, in the next section we optimize each beam for selected target locations behind the blocker, and compare their performance.

\section{Performance assessment of optimized beams}
%

%
%
\begin{figure}[t!]
\centering
    \includegraphics[width=1\linewidth]{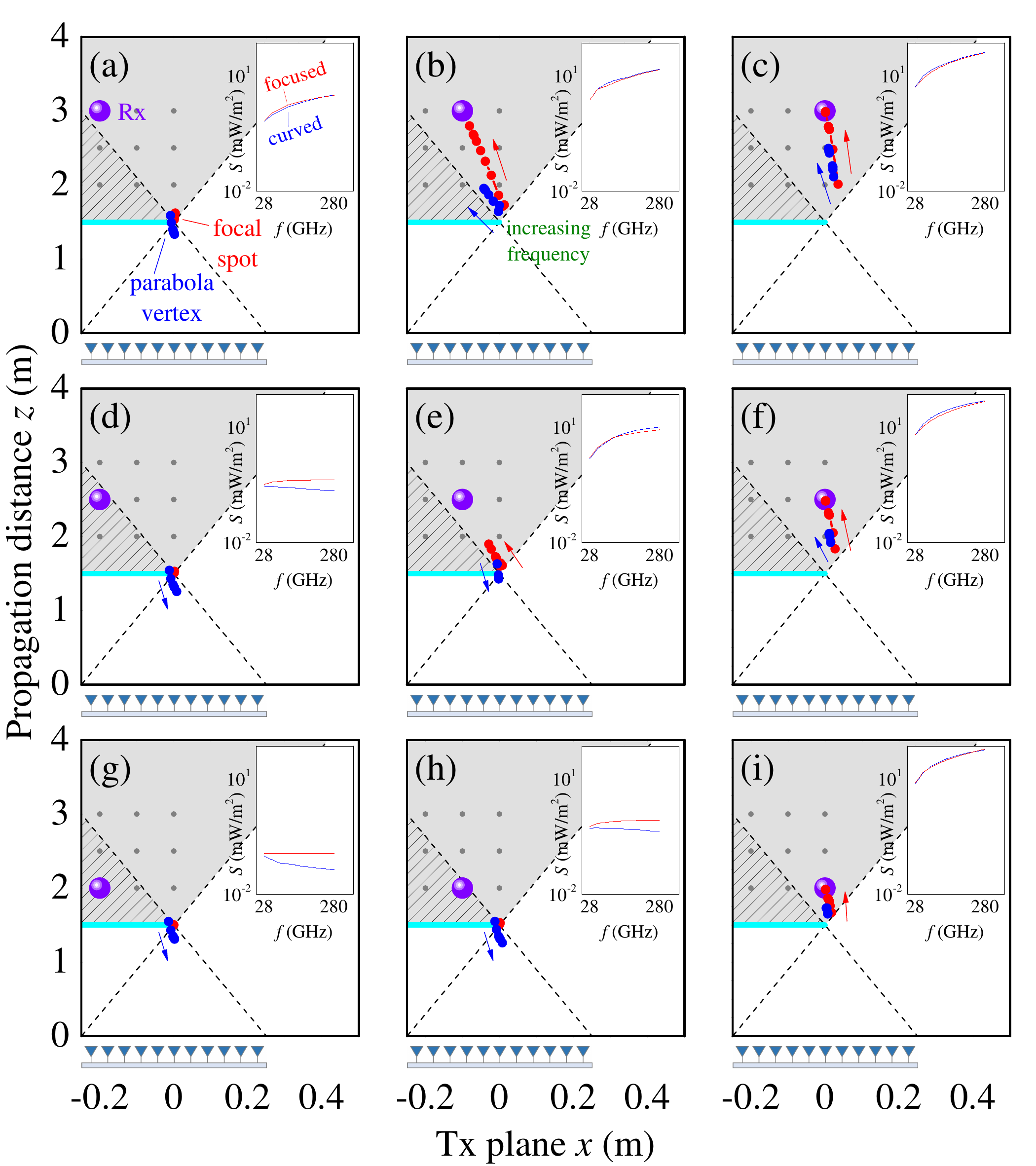}    
	\caption{Performance of optimized curved (blue) and focused (red) beams, for each of the nine selected Rx locations marked in each panel. Each inset shows the optimized power density at each Rx, as a function of the operating frequency, from the FR2 up to the sub-THz spectral range. The partially blocked regime is marked in gray, and the fully blocked regime is denoted with the striped pattern.}
    	\label{fig:fig05}
\end{figure}
%
%

\noindent We select nine target locations for the receiver (Rx) on a $3\times3$ grid that spans the coordinates $x=-0.2\,\mathrm{m},-0.1\,\mathrm{m},0\,\mathrm{m}$ and $z=2\,\mathrm{m},2.5\,\mathrm{m},3\,\mathrm{m}$, marked with the dots shown in the insets of Fig.\,\ref{fig:fig05}. The Rx's that reside in the striped area are fully blocked, and the rest are partially blocked. For each of the nine Rx's we scan the entire parametric space and identify the parameters $x_0,z_0,\beta$ for the curved beam, and the parameters $d_f,\theta_r$ for the focused beam that maximize the power delivery at each Rx. Each panel in Fig.\,\ref{fig:fig05} corresponds to the resulting beam design for each Rx location, with the blue dots denoting the optimized curved beam vertex, and the red dots denoting the location of the optimized focal point, as a function of the operating frequency.  The optimized power density for each Rx is shown in each individual inset in log scale. \\
\indent We observe that, for most of the Rx locations, the performance of both beams is similar throughout the entire spectral range (Figs.\,\ref{fig:fig05}(a),(b),(c),(e),(f),(i)), while for Rx's residing in the fully blocked region (Figs.\,\ref{fig:fig05}(d),(g),(h)) focused beams outperform curved beams at high frequencies; at lower frequencies, where beam formation is inefficient and edge diffraction is more prominent, the performance of both beams becomes again comparable.
Interestingly, for the fully blocked users, the optimized solutions bring the vertex and focal point at the blocker's edge, while when there is some LoS between the user and the UPA, the vertex and focal point move closer to the user; the latter is no surprise, because in the absence of blockage, the power delivery to the user is maximized when the vertex and focal point are exactly at the user location. \\
%
%
%
\begin{figure}[t!]
\centering
    \includegraphics[width=1\linewidth]{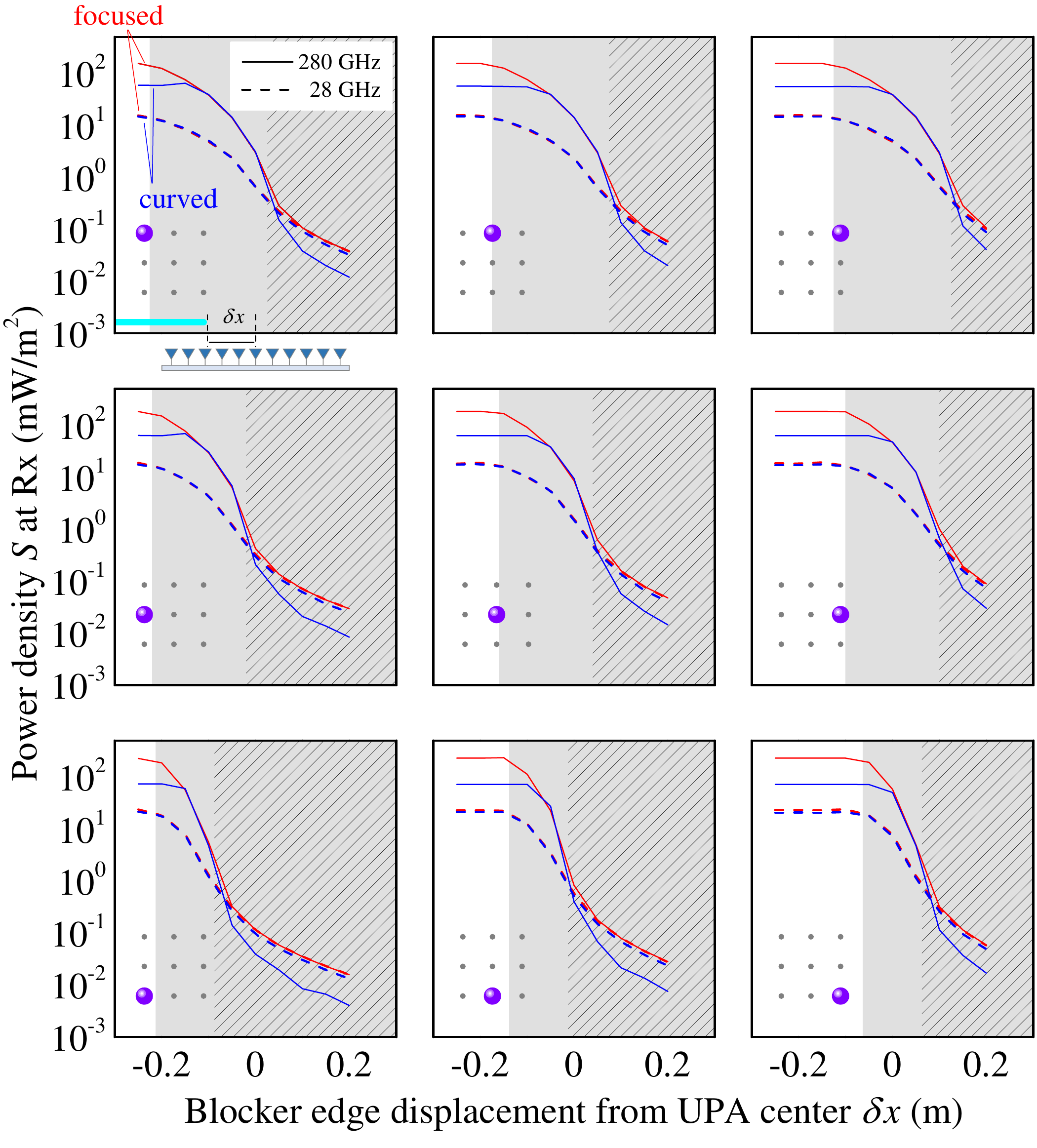}    
	\caption{Performance of optimized curved (blue) and focused (red) beams under variable blockage, for each of the nine Rx locations studied in Fig.\,\ref{fig:fig05}. As the blocker edge displacement $\delta x$ moves towards positive values, the UPA-Rx LoS becomes partially blocked (gray region) and, eventually, entirely blocked (striped region). The displacement $\delta x$ is shown schematically in the top-left panel.}
    	\label{fig:fig06}
\end{figure}
%
%
\indent These results demonstrate that, although the capability of curved beams to propagate on bent trajectories seems the natural choice for propagation behind the LoS, focused beams might be as much efficient for serving users behind the corner. 
In fact, because in LoS scenarios focused beams are able to concentrate all the incident power at the user, they might even be preferable over curved beams. This is investigated in Fig.\,\ref{fig:fig06}, in which we repeat the optimization process for various blockage conditions. From the entire spectral range under study we choose the highest and lowest frequencies ($28\,\mathrm{GHz}$ and $280\,\mathrm{GHz}$, respectively), and we calculate the optimized curved beam and the optimized focused beam, for each of the nine Rx locations, as a function of the blocker displacement from the UPA center, $\delta x$. 
At $28\,\mathrm{GHz}$, where focused beams and curved beams are inefficiently formed for the given Tx aperture, both beams perform similarly, regardless of the blockage conditions. However, at $280\,\mathrm{GHz}$, where beams are formed efficiently, the landscape changes. When $\delta x=-0.2\,\mathrm{m}$, all Rx's are in full LoS with the entire UPA, i.e. they reside in the unblocked region (white area). In this regime, the optimized focused beams outperform the optimized curved beams, as expected. As $\delta x$ gradually increases, the Rx's pass on from the unblocked region to the partially blocked region (gray area), in which both beams perform similarly. Last, for larger $\delta x$, the Rx's become fully blocked, where again the focused beams outperform the curved beams, due to stronger edge diffraction, as previously explained through the examples of Fig.\,\ref{fig:fig04}.

\section{Discussion}

\begin{figure}[t!]
\centering
    \includegraphics[width=1\linewidth]{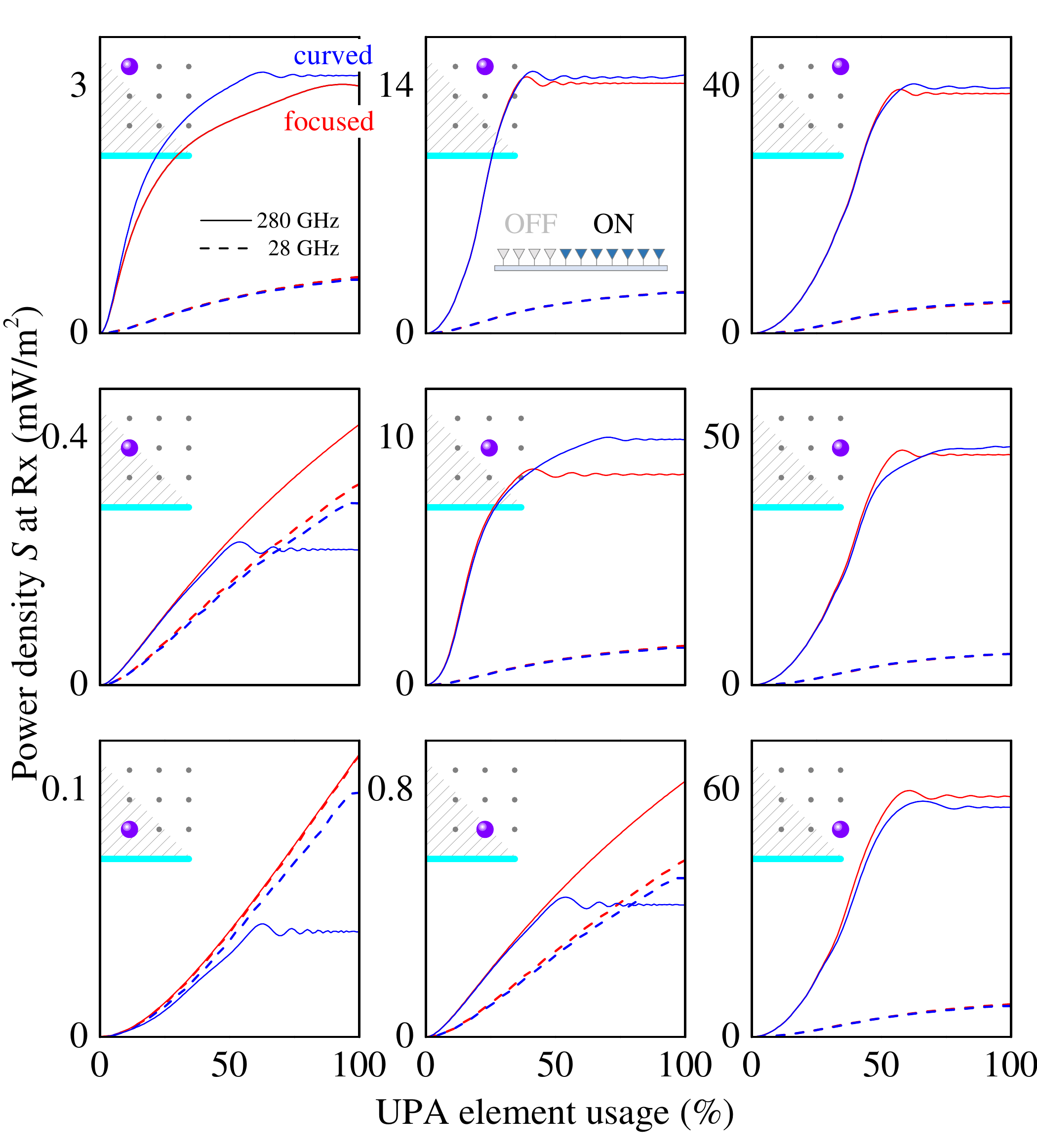}    
	\caption{Performance of the optimized beams of Fig.\,\ref{fig:fig05}, as a function of the operating UPA elements. The element usage is shown schematically in the top-middle panel.}
    	\label{fig:fig07}
\end{figure}

\noindent Despite curved beams being the natural choice in beyond-the-LoS propagation scenarios, our analysis demonstrates that focused beams can perform at least as efficiently. 
The reason for this unexpected and counter-intuitive finding is that the very existence of a curved beam requires the contributions from \textit{all} UPA elements; the contribution from elements that do not have LoS with the user is lost and, hence, the bending capability of the curved beam can be degraded or even lost if a critical number of UPA elements is blocked. In other words, the propagation of curved beams behind a blocker, requires \textit{some} LoS between the UPA and the targeted user location, that is, curved beams cannot propagate in fully blocked regions. Therefore, wavefront engineering alone does not suffice to bend a beam behind the corner; in this case edge diffraction takes over, and the stronger the field at the knife edge, the stronger the field behind the corner, hence the superiority of focused beams.\\
\indent This is further explained in Fig.\,\ref{fig:fig07}, where the performance of the optimized beams of Fig.\,\ref{fig:fig05} is investigated as a function of the UPA element usage. 
That is, for each of the nine Rx locations we choose the optimal curved beam and the optimal focused beam that we calculated earlier, and starting from $x=-L_x/2$, we gradually disable the elements of the UPA so that we first eliminate those that are already blocked. 
In the panels of Fig.\,\ref{fig:fig07} this action corresponds to moving leftwards from $100\%$ (all elements active) to $0\%$ (all elements disabled). At $280\,\mathrm{GHz}$, where beam formation is most efficient, the performance of curved beams practically does not change until all elements that are already blocked have been switched off ($\approx 50\%$). Only then does the power density start decreasing, because beyond this point we are disabling elements that are unblocked and therefore contribute significantly to the beam. 
For the focused beam, the picture is slightly different: for users that are partially blocked wavefront engineering takes over, and because the focal point is directed at each user's location, the elements that are already blocked play little or no role, as with the curved beam. However, for users that are fully blocked, edge diffraction takes over, and because the focal point is directed at the blocker's edge, lower element usage results in less power at the focal point and consequently behind the blocker.
At $28\,\mathrm{GHz}$, where beam formation is inefficient, the performance of both beams is dominated by edge diffraction, regardless of the extent of blockage. In this case, because both the optimized parabola vertex and the focal point are located close to the blocker's edge, reducing the element usage decreases the power at the parabola vertex and the focal point and consequently behind the blocker. \\
%
%
%
%
%
%
\indent When the Tx aperture is not sufficiently electrically large, and beam formation is inefficient, both beams perform similarly, regardless of the blockage conditions. For example, at $28\,\mathrm{GHz}$, where inefficient beam formation is observed, the aperture is only $L_x\approx 46\lambda$. Note that, this is not a limitation per se, as for larger aperture, beam formation can be efficient at this frequency as well. The limitation is more on the practicality, as with decreasing frequency the wavelength increases, requiring very large -possibly impractically large- arrays. \\
\indent In general, when the Tx aperture is electrically large, so that beam formation is efficient, focused beams outperform curved beams in the unblocked and fully blocked UPA regimes, while performing similarly in the partially blocked UPA regime. In the unblocked and partially blocked regime, on the one hand, focused beams concentrate all the incident power on the user, hence they are advantageous over curved beams by design. In the fully blocked region, on the other hand, both beams are targeted at the blocker's edge; propagation beyond LoS in this case is dominated by edge diffraction, hence focused beams perform better, because they supply the diffracted beam with more power than curved beams. Of course, because focused beams concentrate the transmitted power at a very limited area in space, their superiority could be compromised by their vulnerability to subtle misplacements of the focal point from the target location. In this sense, curved beams could possibly turn out to be more efficient in certain scenarios.
\section{Conclusion}
In this work, we explained the three basic mechanisms of beam propagation beyond the LoS, namely RI modulation, edge diffraction, and wavefront engineering. 
We demonstrated that, while pure beam bending can be achieved through the RI modulation of the medium in which the beam evolves, this technique is rather challenging for free-space propagation, due to the weak temperature dependence of the air RI. 
Therefore, beam bending behind corners usually results from an interplay between wavefront engineering and edge diffraction.
We identified three distinct regimes, namely unblocked, partially blocked, and fully blocked, and we showed that beam bending through wavefront engineering dominates in the unblocked and partially blocked regimes, while edge diffraction dominates in the fully blocked regime; as a result, curved beams cannot really bend behind the corner, unless there is some LoS between the user and the transmitter. Based on our findings, we employed focused beams and we demonstrated that they perform at least as efficiently as curved beams.

\section*{Acknowledgments}
This work was supported by the European Commission’s Horizon Europe Programme under the Smart Networks and Services Joint Undertaking INSTINCT project (Grant Agreement $101139161$).

\section*{APPENDIX}
\subsection{Angular spectrum representation}
\label{Sec:AppendixA}

To numerically propagate the beams, we use the angular spectrum representation, which implements the exact scalar Helmholtz propagation on a sampled grid including evanescent components. According to this technique, the propagation from $z$ to $z+\delta z$ is expressed as \cite{Droulias2022, Droulias2024a}
\begin{equation}
    E(x,y,z+\delta z) = \mathcal{FT}\,^{-1}\{\,\mathcal{FT}\,[E(x,y,z)]e^{jk_z\delta z}\},
    \label{Eq:EqX02}
\end{equation}
where \(\mathcal{FT}\) denotes the Fourier transform and \(\mathcal{FT}\)\,$^{-1}$ its inverse. First, we discretize the entire propagation distance $z_\mathrm{total}$ in $N_z$ segments of length $\delta z$, so that $z_\mathrm{total}=N_z \delta z$. At $z=0$ we use the input beam profile \eqref{Eq:EqINPUTWAVE} to calculate the beam for the first $\delta z$ segment. Then, we use the output beam as input, and propagate for the next $\delta z$ and so on, until we propagate the beam for the entire distance. 
In our examples $z_\mathrm{total}=4\,\mathrm{m}$ and $N_z=200$. 
In the 1D beam examples, we use a 1D transverse computational grid of size  $N_x=4096$ and $\delta x=\lambda/2$. In the 2D beam examples, we use a 2D transverse computational grid of size $N_x \times N_y$ with $N_x=N_y=4096$ and $\delta x=\delta y=\lambda/2$. To absorb the energy that possibly spreads beyond the extent of the grid, we select absorptive boundaries and apply a super-Gaussian filter function. 

\subsection{Huygens-Fresnel integral}
\label{Sec:AppendixB}

For the analytical calculations, we use the Huygens-Fresnel integral under the Fresnel approximation, which is the paraxial approximation to the exact angular spectrum representation, and is therefore suitable for the examples considered in this work. 
The Huygens-Fresnel integral is expressed as
\begin{align}
    \label{Eq:EqAPP_HF}
    E(\mathbf{r})=-\frac{ik}{2\pi}\iint A(\mathbf{r}')e^{j\phi(\mathbf{r}')}\frac{\exp{\left(-ik|\mathbf{r}-\mathbf{r}'| \right)}}{|\mathbf{r}-\mathbf{r}'|} d\mathbf{r}',
\end{align}
where $E(\mathbf{r})$ is the $E$-field at $\mathbf{r}=(x,y,z)$, $\mathbf{r}'=(x',y',0)$, $d\mathbf{r}'=dx'dy'$, and the integration is performed within the UPA surface, i.e. $|x'|<L_x/2$ and $|y'|<L_y/2$.
The term $A(\mathbf{r}')e^{j\phi(\mathbf{r}')}$ in the integrand is the UPA input beam profile, which is given by \eqref{Eq:EqINPUTWAVE}. To perform the integration, we first simplify the Green's function in the integrand, by expanding $|\mathbf{r}-\mathbf{r}'|$ and keeping the first few terms, i.e. we write
\begin{align}
    \label{Eq:EqAPP_FRapprx}
    |\mathbf{r}-\mathbf{r}'|=\sqrt{(x-x')^2+(y-y')^2+z^2}\\
    \approx z+\frac{(x-x')^2+(y-y')^2}{2z}
    \nonumber
\end{align}
This approximation is also known as the Fresnel approximation and is valid for $\lambda\ll|\mathbf{r}-\mathbf{r}'|\ll z$, as in our case. For the $1/|\mathbf{r}-\mathbf{r}'|$ term of the Green's function it suffices to keep only the $z$ term, and \eqref{Eq:EqAPP_HF} becomes
\begin{align}
    \label{Eq:EqAPP_HFapprx}
    & E(x,y,z) =-\frac{ik e^{-ik z}}{2\pi z} \times \\
    & \int_{-L_y/2}^{+L_y/2} \int_{-L_x/2}^{+L_x/2} A(x')e^{j\phi(x')}e^{\frac{-ik}{2z}\left[(x-x')^2+(y-y')^2\right] } dx' dy'.
    \nonumber
\end{align}
In this work we have used uniform amplitude for the input beam, hence $A(x)=A_0$ in \eqref{Eq:EqHFintegral}.

\subsection{Power density approximations}
\label{Sec:AppendixC}

\subsubsection{Edge diffraction}
%
%
\begin{figure}[t!]
\centering
    \includegraphics[width=1\linewidth]{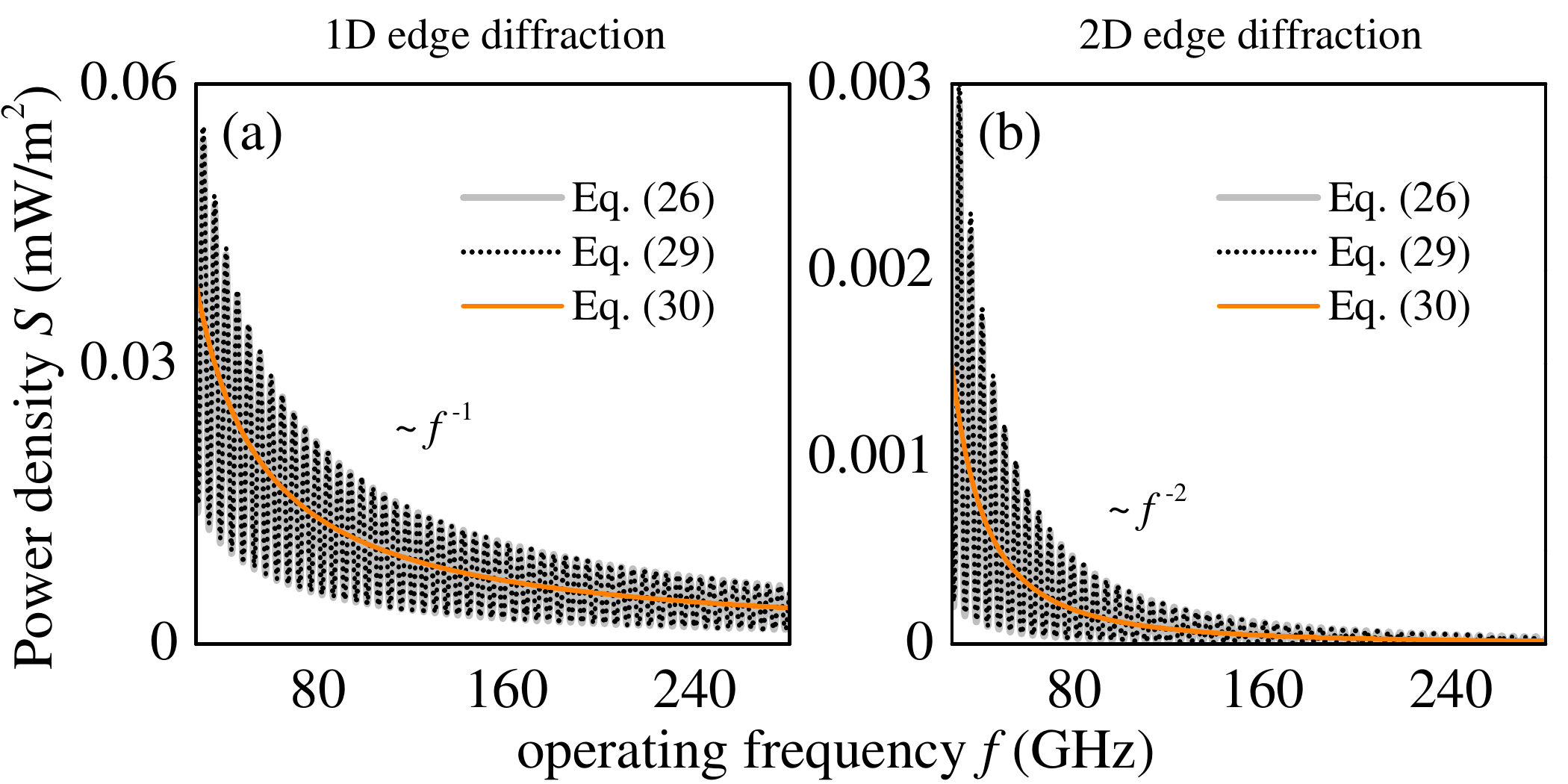}    
	\caption{Efficiency of edge diffraction, as a function of the operating frequency. (a) $S_\mathrm{diff,1D}$ at $(x, y, z) = (-0.5\,\mathrm{m},0\,\mathrm{m},4\,\mathrm{m})$. (b) $S_\mathrm{diff,2D}$ at $(x, y, z) = (-0.5\,\mathrm{m},0.5\,\mathrm{m},4\,\mathrm{m})$. The beam is generated at $z=0$ with $\phi(x)=0$.}
    	\label{fig:figAPP01}
\end{figure}
%

The power density \eqref{Eq:EqSdiff} can be concisely written as
\begin{align}
    \label{Eq:EqAPPSdiff2D}
    S_\mathrm{diff,2D} =  \frac{A_0^2}{2\eta}R(L_x,x)R(L_y,y),
\end{align}
where
\begin{align}
    \label{Eq:EqAPPR}
    R(u,v)= \left|\frac{\mathrm{erf} (q(u,v))+\mathrm{erf}(q(u,-v))}{2}\right|^2,
\end{align}
and the argument $q$ is given by
\begin{align}
    q(u,v) = \left(\frac{1}{4}+\frac{i}{4} \right) \sqrt{\frac{k}{z}}(u+2 v).
\end{align}
For 1D diffraction, $u\rightarrow\infty$ leads to $R(u,v)\rightarrow 1$, hence 
\begin{align}
    \label{Eq:EqAPPSdiff1D}
    S_\mathrm{diff,1D} =  \frac{A_0^2}{2\eta}R(L_x,x),
\end{align}
where we have used $u=L_y$, as in our examples.
For $k\gg (u+2v)^2/z$, $|v|>u/2$, we can approximate $R(u,v)$ as
\begin{align}
    \label{Eq:EqAPPRapprx1}
    R(u,v) \approx \frac{4z(u^2+4 v^2)}{\pi(u^2-4 v^2)^2} \left(1+ \frac{u^2-4 v^2}{u^2+4 v^2} \cos{\left(\frac{k u v}{z}\right)} \right)\frac{1}{k}.
\end{align}
Eliminating the oscillatory term leads to
\begin{align}
    \label{Eq:EqAPPRapprx2}
    R(u,v) \approx \frac{4z(u^2+4 v^2)}{\pi(u^2-4 v^2)^2}\frac{1}{k}.
\end{align}
In Fig.\,\ref{fig:figAPP01}(a) we plot \eqref{Eq:EqAPPSdiff1D} at the observation point $(x, y, z) = (-0.5\,\mathrm{m},0\,\mathrm{m},4\,\mathrm{m})$, using the full expression \eqref{Eq:EqAPPR}, as well as the approximate forms \eqref{Eq:EqAPPRapprx1} and \eqref{Eq:EqAPPRapprx2}. The parameters used in this example are $L_x=0.5\,\mathrm{m}$, $A_0=\sqrt{2\eta P_t/L_x}$ and $P_t=1\,\mathrm{mW}$. \\
\indent In Fig.\,\ref{fig:figAPP01}(b) we plot \eqref{Eq:EqAPPSdiff2D} at the observation point $(x, y, z) = (-0.5\,\mathrm{m},0.5\,\mathrm{m},4\,\mathrm{m})$, using the full expression \eqref{Eq:EqAPPR}, as well as the approximate forms \eqref{Eq:EqAPPRapprx1} and \eqref{Eq:EqAPPRapprx2}. The parameters used in this example are $L_x=L_y=0.5\,\mathrm{m}$, $A_0=\sqrt{2\eta P_t/L_xL_y}$ and $P_t=1\,\mathrm{mW}$. \\
\indent For both 1D and 2D diffraction, the approximation \eqref{Eq:EqAPPRapprx1} successfully reproduces the full expression with \eqref{Eq:EqAPPR}. The approximation \eqref{Eq:EqAPPRapprx2} illustrates how the diffraction efficiency changes as $\sim f^{-1}$ in the 1D case and as $\sim f^{-2}$ in the 2D case.

\subsubsection{Focused beams}

%
%
\begin{figure}[t!]
\centering
    \includegraphics[width=1\linewidth]{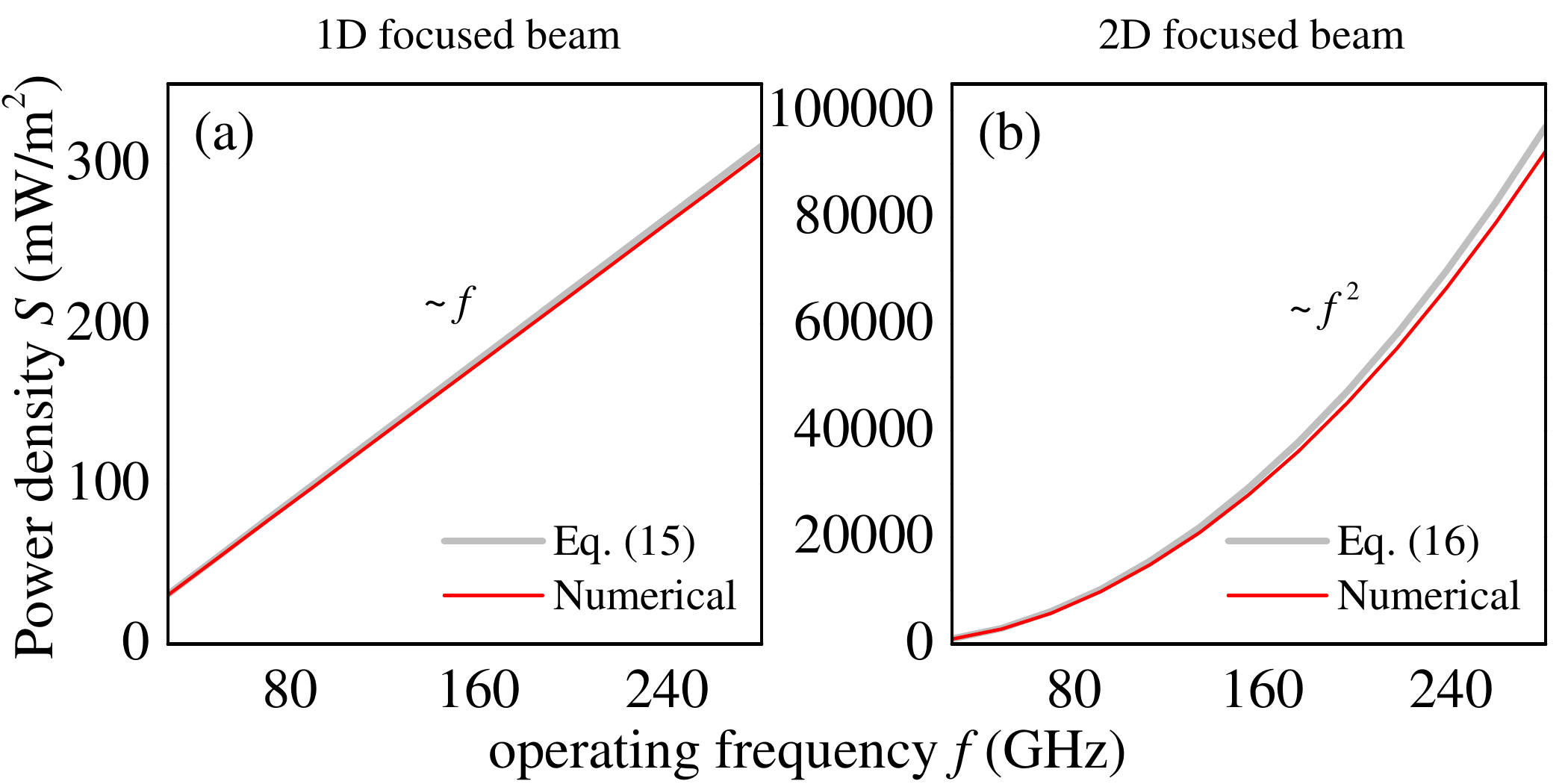}    
	\caption{Efficiency of focused beam formation, as a function of the operating frequency. (a) $S_\mathrm{foc,1D}$. (b) $S_\mathrm{foc,2D}$.}
    	\label{fig:figAPP02}
\end{figure}
%

To verify our analytical results, we numerically propagate a 1D focused beam with parameters $\theta_r=0^\circ$, $d_f=1.5\,\mathrm{m}$, using a UPA with $L_x = 0.5\,\mathrm{m}$. For the beam excitation we use $A_0=\sqrt{2\eta P_t/L_x}$ with $P_t=1\,\mathrm{mW}$ and $\phi(x)=kd_f+k x^2/2d_f$. In Fig.\,\ref{fig:figAPP02}(a) we plot the numerically calculated power density at the focal point $(x, y, z) = (0\,\mathrm{m},0\,\mathrm{m},1.5\,\mathrm{m})$, together with the analytical expression \eqref{Eq:EqSfoc1D}. \\
\indent Next, we use a UPA of size $L_x = L_y = 0.5\,\mathrm{m}$ and propagate a 2D focused beam towards the same focal point. For the beam excitation we use $A_0=\sqrt{2\eta P_t/L_xL_y}$ with $P_t=1\,\mathrm{mW}$ and $\phi(x)= kd_f + k (x^2+y^2)/2d_f$. In Fig.\,\ref{fig:figAPP02}(b) we plot the numerically calculated power density at the focal point $(x, y, z) = (0\,\mathrm{m},0\,\mathrm{m},1.5\,\mathrm{m})$, together with the analytical expression \eqref{Eq:EqSfoc2D}. \\
\indent Our simulations verify that the focused beam efficiency changes as $\sim f$ in the 1D case and as $\sim f^2$ in the 2D case. The slight deviation between the analytical expressions and the numerics is due to the Fresnel approximation \eqref{Eq:EqAPP_FRapprx} that has been used to simplify the integrations in the analytical derivations.
\subsubsection{Curved beams}

\begin{figure}[t!]
\centering
    \includegraphics[width=1\linewidth]{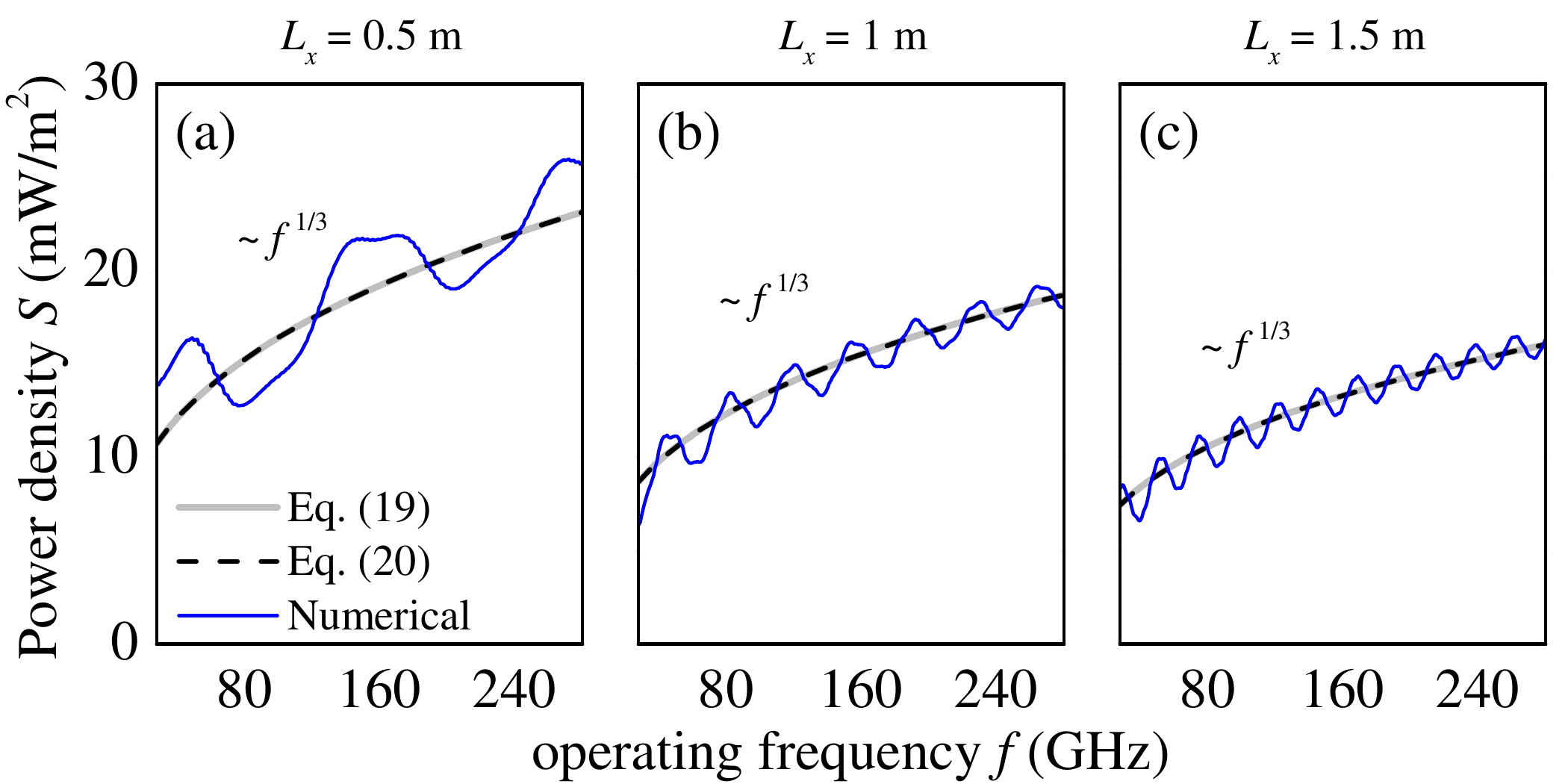}    
	\caption{Efficiency of curved beam formation, as a function of the operating frequency, for different UPA sizes. (a) $L_x=0.5\,\mathrm{m}$. (b) $L_x=1\,\mathrm{m}$. (c) $L_x=1.5\,\mathrm{m}$.}
    	\label{fig:figAPP03}
\end{figure}

To verify our analytical results, we numerically propagate a curved beam with parameters $x_0=0\,\mathrm{m}$, $z_0=1.5\,\mathrm{m}$ and $\beta = -0.11\,\mathrm{m^{-1}}$, using a UPA with $L_x = 0.5\,\mathrm{m}$. For the beam excitation we use $A_0=\sqrt{2\eta P_t/L_x}$ with $P_t=1\,\mathrm{mW}$. In Fig.\,\ref{fig:figAPP03}(a) we plot the numerically calculated power density at the parabola vertex $(x, y, z) = (0\,\mathrm{m},0\,\mathrm{m},1.5\,\mathrm{m})$, together with the analytical expressions \eqref{Eq:EqSAiry1D_at0}, \eqref{Eq:EqSAiry1D}, using $\alpha = 5.4\,\mathrm{m^{-1}}$. The performance of the numerically propagated beam has an oscillatory evolution with operating frequency, which nevertheless increases similarly to the theoretical Airy beam. The qualitative difference between the two is due to the different excitation conditions for the two beams; while both beams are designed to evolve on the same parabolic path, the Airy beam excitation involves both specific amplitude and phase distributions, while in our simplified version we use uniform amplitude. \\
\indent In Fig.\,\ref{fig:figAPP03}(b) we increase the UPA size to $L_x = 1\,\mathrm{m}$, and we use $\beta = -0.22\,\mathrm{m^{-1}}$ to adjust the parabola vertex to the same location as in the previous example. In the analytical expressions we use $\alpha = 2.8\,\mathrm{m^{-1}}$. In Fig.\,\ref{fig:figAPP03}(c) we repeat the calculations using $L_x = 1.5\,\mathrm{m}$, $\beta = -0.33\,\mathrm{m^{-1}}$ and $\alpha = 1.8\,\mathrm{m^{-1}}$. We observe that, with increasing UPA size, the performance of the numerically propagated beam converged to that of the analytical Airy beam, cofirming the $\sim f^{1/3}$ performance.
%

%
%
%
%

\bibliographystyle{IEEEtran}
\bibliography{IEEEabrv,main}

\vfill

\end{document}